\documentclass[twocolumn,floats,prl,aps,unsortedaddress,superscriptaddress]{revtex4-1}

\usepackage[dvips]{graphicx}
\usepackage{amsfonts}
\usepackage{amsmath}
\usepackage{amssymb}
\usepackage{exscale}
\usepackage{color}	
\usepackage{braket}	
\usepackage{multirow}	
\usepackage{epstopdf}

\usepackage[normalem]{ulem}    

\begin{document}
\title{Engineering two-dimensional electron gases 
	   at the $(001)$ and $(101)$ surfaces of TiO$_2$ anatase using light}

\author{T.~C.~R\"odel}
\affiliation{Centre de Sciences Nucl\'eaires et de Sciences de la Mati\`ere (CSNSM), 
	Universit\'e Paris-Sud and CNRS/IN2P3, B\^atiments 104 et 108, 91405 Orsay cedex, France}
\affiliation{Synchrotron SOLEIL, L'Orme des Merisiers, Saint-Aubin-BP48, 91192 Gif-sur-Yvette, France}
\author{F.~Fortuna}
\affiliation{Centre de Sciences Nucl\'eaires et de Sciences de la Mati\`ere (CSNSM), 
	Universit\'e Paris-Sud and CNRS/IN2P3, B\^atiments 104 et 108, 91405 Orsay cedex, France}
\author{F.~Bertran}
\affiliation{Synchrotron SOLEIL, L'Orme des Merisiers, Saint-Aubin-BP48, 91192 Gif-sur-Yvette, France}
\author{M.~Gabay}
\affiliation{Laboratoire de Physique des Solides, Universit\'e Paris-Sud, B\^atiment 510, 91405 Orsay, France}
\author{M.~J.~Rozenberg}
\affiliation{Laboratoire de Physique des Solides, Universit\'e Paris-Sud, B\^atiment 510, 91405 Orsay, France}
\author{A.~F.~Santander-Syro}
\email{andres.santander@csnsm.in2p3.fr}
\affiliation{Centre de Sciences Nucl\'eaires et de Sciences de la Mati\`ere (CSNSM), 
	Universit\'e Paris-Sud and CNRS/IN2P3, B\^atiments 104 et 108, 91405 Orsay cedex, France}
\author{P.~Le~F\`evre}
\affiliation{Synchrotron SOLEIL, L'Orme des Merisiers, Saint-Aubin-BP48, 91192 Gif-sur-Yvette, France}

\begin{abstract}
	We report the existence of metallic two dimensional electron gases (2DEGs) 
	at the $(001)$ and $(101)$  surfaces of bulk-insulating TiO$_2$ anatase 
	due to local chemical doping by oxygen vacancies in the near-surface region.
	Using angle-resolved photoemission spectroscopy,  
	we find that the electronic structure at both surfaces is composed 
	of two occupied subbands of $d_{xy}$ orbital character. 
	While the Fermi surface observed at the $(001)$ termination is isotropic, 
	the 2DEG at the $(101)$ termination is anisotropic and shows a charge carrier density
	three times larger than at the $(001)$ surface. 
	Moreover, we demonstrate that intense UV synchrotron radiation 
	can alter the electronic structure and stoichiometry of the surface up to the 
	complete disappearance of the 2DEG. 
	These results open a route for the nano-engineering of confined electronic states,
	the control of their metallic or insulating nature, and the tailoring of their 
	microscopic symmetry, using UV illumination at different surfaces of anatase.
\end{abstract}
\maketitle

In its pure stoichiometric form, the transition metal oxide (TMO) TiO$_2$ 
is a transparent insulator that crystallizes in mainly two different phases: 
rutile and anatase.
Both phases have been studied extensively over the last decades,
due to their photocatalytic properties 
discussed in several reviews~\cite{Diebold2003,Zhang2012,Henderson2011,Fujishima2008}.
Recently, a strong interest in the anatase phase of TiO$_2$ 
also surged, owing to its potential for applications in other research fields.
For instance, networks of anatase nanoparticles 
are found in dye-sensitized solar cells~\cite{Hagfeldt2010,Gratzel2009}, 
anatase thin films can be used as transparent conducting oxides~\cite{Furubayashi2005}, 
and devices based on anatase
can be envisioned in spintronics~\cite{Matsumoto2001,Fukumura2004}. 
To harness such a wide range of functionalities and guide potential applications using anatase,
it is thus critical to understand its microscopic electronic structure, 
which will be ultimately responsible for the remarkable properties of this material. 
Moreover, as most applications in microelectronics or heterogeneous catalysis 
involve essentially the electronic states at the material's surface,
it is crucial to directly measure and characterize such states. 

More generally, the study of two-dimensional electron gases (2DEGs) 
in TMO surfaces/interfaces has become a very active field of research. 
The archetypal example, SrTiO$_3$-based heterostructures, 
display many fundamentally interesting properties~\cite{Mannhart2010,Zubko2011,Hilgenkamp2013},
such as field-effect induced insulator-to-superconductor transitions~\cite{Caviglia2008}, 
magnetism~\cite{Brinkman2007} and the coexistence of magnetism and superconductivity~\cite{Dikin2011}.
More recently, the discoveries of 2DEGs at the bare $(001)$, $(110)$ and $(111)$ surfaces 
of SrTiO$_3$~\cite{Santander-Syro2011,Meevasana2011, Plumb2013, Wang2014,Roedel2014,Walker2014} 
and KTaO$_3$~\cite{King2012,Santander-Syro2012,Bareille2014},
triggered new avenues of research.
Additionally, very recent spin-resolved measurements of the 2DEG on SrTiO$_3$~$(001)$
revealed a giant spin splitting of its electronic structure~\cite{Santander-Syro2014},
epitomizing the richness of physical properties that can be found in these 2DEGs.

It is well established that the TiO atomic planes, 
and their ability to accommodate chemical doping by oxygen vacancies at the surface region, 
play a key role in the formation of the 2DEG at the surface of SrTiO$_3$~$(001)$. 
Thus, as a step forward to understand the formation of 2DEGs in TMOs, 
it is natural to focus on pure TiO$_2$ crystals such as rutile or anatase.
In fact, it is known that, for both rutile and anatase crystal surfaces, 
UV or electron irradiation creates oxygen vacancies~\cite{Knotek1978,Shultz1995,Tanaka2004}. 
But while 2DEGs can be readily obtained at various anatase surfaces, as we will show further, 
the failure to observe 2DEGs on the cleaved or \emph{in-situ} 
prepared surfaces of \emph{rutile} TiO$_2$~$(110)$
(Ref.~\cite{Moser2013} and Supplemental Material),
indicates that other factors are equally important.
In fact, excess electrons in TiO$_2$, due to oxygen vacancies at the surface, 
form polarons which behave quite differently for the rutile and anatase phases~\cite{Setvin2014}. 
This demonstrates that structural factors, such as surface lattice distorsions,
can determine the fate of those electrons.
Thus, to elucidate the origin and realization of 2DEGs in different oxides, 
it is essential to probe the surface electronic structure arising in different lattice configurations.

Here we report the observation of 2DEGs at the $(001)$ and $(101)$ surfaces 
of TiO$_2$ anatase. Using angle resolved photoemission spectroscopy (ARPES),
we find that, for both surfaces, the electronic structure consists of two light bands 
of $d_{xy}$ character. While the Fermi surface at the $(001)$ surface is isotropic, 
the one at the $(101)$ surface is highly anisotropic, 
reflecting the symmetry of the surface layer~\cite{Roedel2014}. 
Moreover, we show that the character of the excess electrons (localized or delocalized) 
depends on the concentration of oxygen vacancies, and that the 2DEG disappears,
leaving no mobile conduction electrons near the Fermi level,
if the surface is reduced too much by intense UV irradiation.

\begin{figure*}
  \begin{minipage}[c]{0.63\textwidth}
  \begin{center}
    \includegraphics[clip, width=\textwidth]{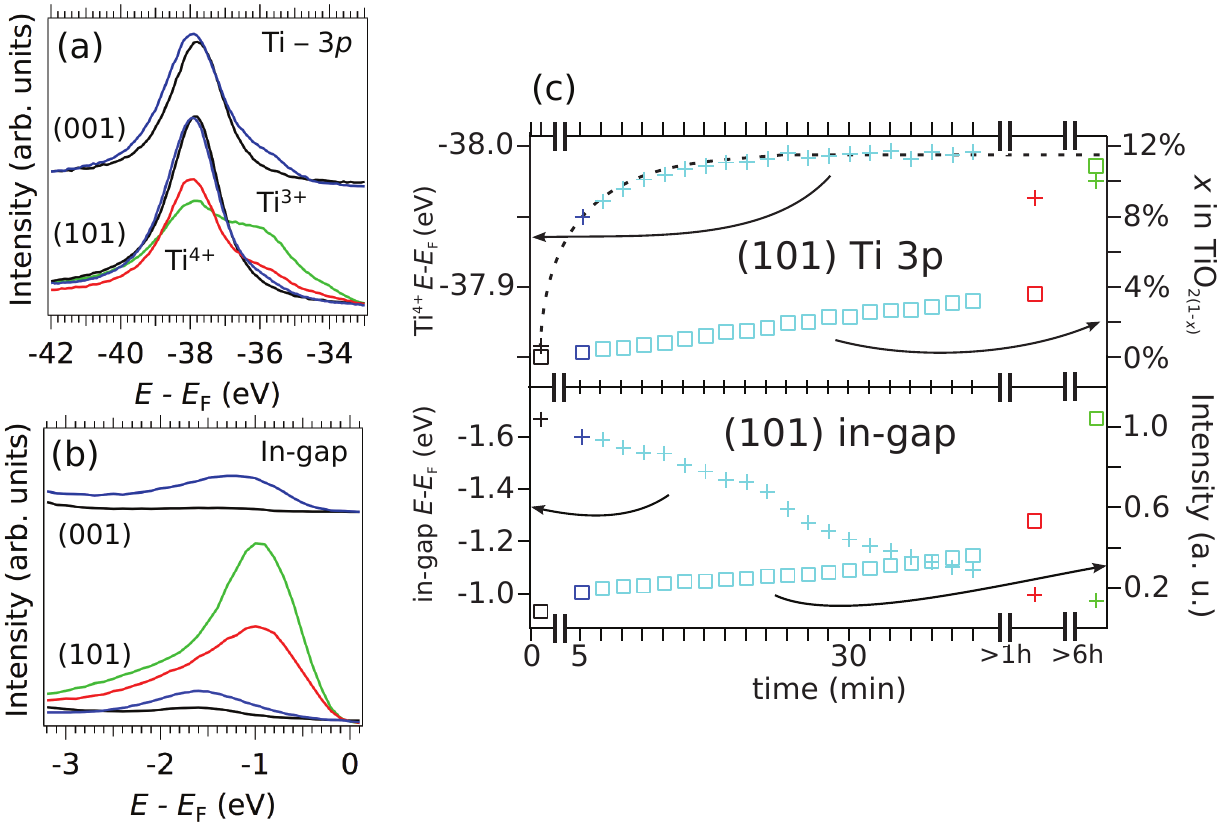}
  \end{center}
  \end{minipage}\hfill
  \begin{minipage}[c]{0.32\textwidth}
    \caption{\label{fig:anatase_UPS} \footnotesize{(Color online) 
  		Angle integrated spectra of (a)~the Ti-$3p$ peak, and (b)~the in-gap state 
  		of the anatase TiO$_2$~$(001)$ and $(101)$ surfaces (upper and lower curves, respectively)
  		measured at $h\nu=100$~eV. 
  		The black curve was measured shortly after the first light exposure, 
  		and the blue, red and green curves at sequentially increasing times.
  		(c)~Binding energy of the Ti$^{4+}$ peak and concentration of oxygen vacancies 
  		(ratio between the Ti$^{3+}$ and total Ti-$3p$ peak area)
  		for different durations of irradiation exposure of the $(101)$ surface (upper panel),
  		and binding energy and intensity of the in-gap state (lower panel).
  		Analogous results, not shown, are obtained for the $(001)$ surface
  		}
  }
  \end{minipage}
\end{figure*}

The ARPES measurements were conducted at the Synchrotron Radiation Center 
(SRC, University of Wisconsin, Madison) and the CASSIOPEE beamline of Synchrotron Soleil (France)
on surfaces cleaved at low temperatures under ultra-high vacuum
-- see Supplemental Material for details about the sample preparation and measurements.
All through this paper, directions and planes are defined 
in the simple tetragonal conventional cell of anatase.
We note $[hkl]$ the crystallographic directions in real space, 
$\langle hkl \rangle$ the corresponding  directions in reciprocal space, 
and $(hkl)$ the planes orthogonal to those directions. 
The indices $h$, $k$, and $l$ of $\Gamma_{hkl}$ correspond to
the reciprocal lattice vectors of the body-centered unit cell.

As shown in Figs.~\ref{fig:anatase_UPS}(a) and ~\ref{fig:anatase_UPS}(b),
the presence of oxygen vacancies in the surface region of anatase can be identified
in the photoemission spectra by the formation of a shoulder in the Ti-$3p$ peak
related to a lower valence of the Ti cation, and the appearance
of an in-gap state~\cite{Thomas2003} around $1$~eV below the Fermi level ($E_F$), 
corresponding to electrons trapped near oxygen vacancies~\cite{Setvin2014}, 
The black curves were measured shortly after the first exposure of the sample 
to the synchrotron radiation, while the blue, red and green curves were recorded
at later subsequent times specified respectively by the abscissas of the blue, red and green
open squares in Fig.~\ref{fig:anatase_UPS}(c). 

As seen in Fig.\ref{fig:anatase_UPS}(a), the peak position of the Ti-$3p$ peak 
shifts to higher binding energies upon UV irradiation, 
demonstrating the band bending (bb) at the surface. 
We fit the  Ti-$3p$ peak by one or two Voigt functions plus a Shirley background 
to determine the contribution of Ti$^{4+}$ and Ti$^{3+}$ states to the line shape. 
This yields a shift of Ti$^{4+}$ peak of $E_{bb}^{(001)}=110$~meV for the $(001)$ surface, 
and $E_{bb}^{(101)}=150$~meV for the $(101)$ surface, 
which corresponds to the \emph{minimal} band bending at the surfaces
averaged over the photoemission probing depth at the used photon energy. 
The actual band bending might be larger, as oxygen vacancies might be already induced 
at the surface by the cleaving process or by the short beam exposure before the first measurement.

The red and green curves in Figs.~\ref{fig:anatase_UPS}(a) and~\ref{fig:anatase_UPS}(b), 
show that many more oxygen vacancies can be induced at the TiO$_2$~$(101)$ surface 
than at the $(001)$ surface, 
in agreement with previous studies~\cite{Thomas2007}.
The band bending (peak position of the Ti$^{4+}$ peak) and concentration of oxygen vacancies 
(ratio between the Ti$^{3+}$ and total Ti-$3p$ peak area)
for the blue, green, and red curves 
as well as for intermediate measurements are plotted in figure~\ref{fig:anatase_UPS}(c)
for the specific case of the anatase $(101)$ surface. 
This shows that the band bending 
saturates at rather low concentration of vacancies at the surface. 
In fact, recent theoretical studies on the electronic structure of SrTiO$_3$~\cite{Jeschke2014,Hao2014} 
suggest that, for a high concentration of oxygen vacancies, 
excess electrons do not fill up the conduction band any longer,
but form only localized states, and thus do not contribute to the rigid band shift.
Additionally, as we will see later, the 2DEG disappears 
if a high concentration of oxygen vacancies is created.
	     
Similarly, as shown in figures~\ref{fig:anatase_UPS}(b) and (c) (lower panel),
the intensity and binding energy of the in-gap state evolve significantly 
upon UV irradiation, both for the $(001)$ and $(101)$ surfaces. 
Possible explanations for the different binding energies are that the oxygen vacancies 
are located at different lattice sites and/or forming clusters~\cite{Jeschke2014}, 
or the existence of different types of defects~\cite{Valentin2009,Sanchez-Sanchez2013}.

\begin{figure*}[t]
  \begin{center}
   	  \includegraphics[clip, width=16cm]{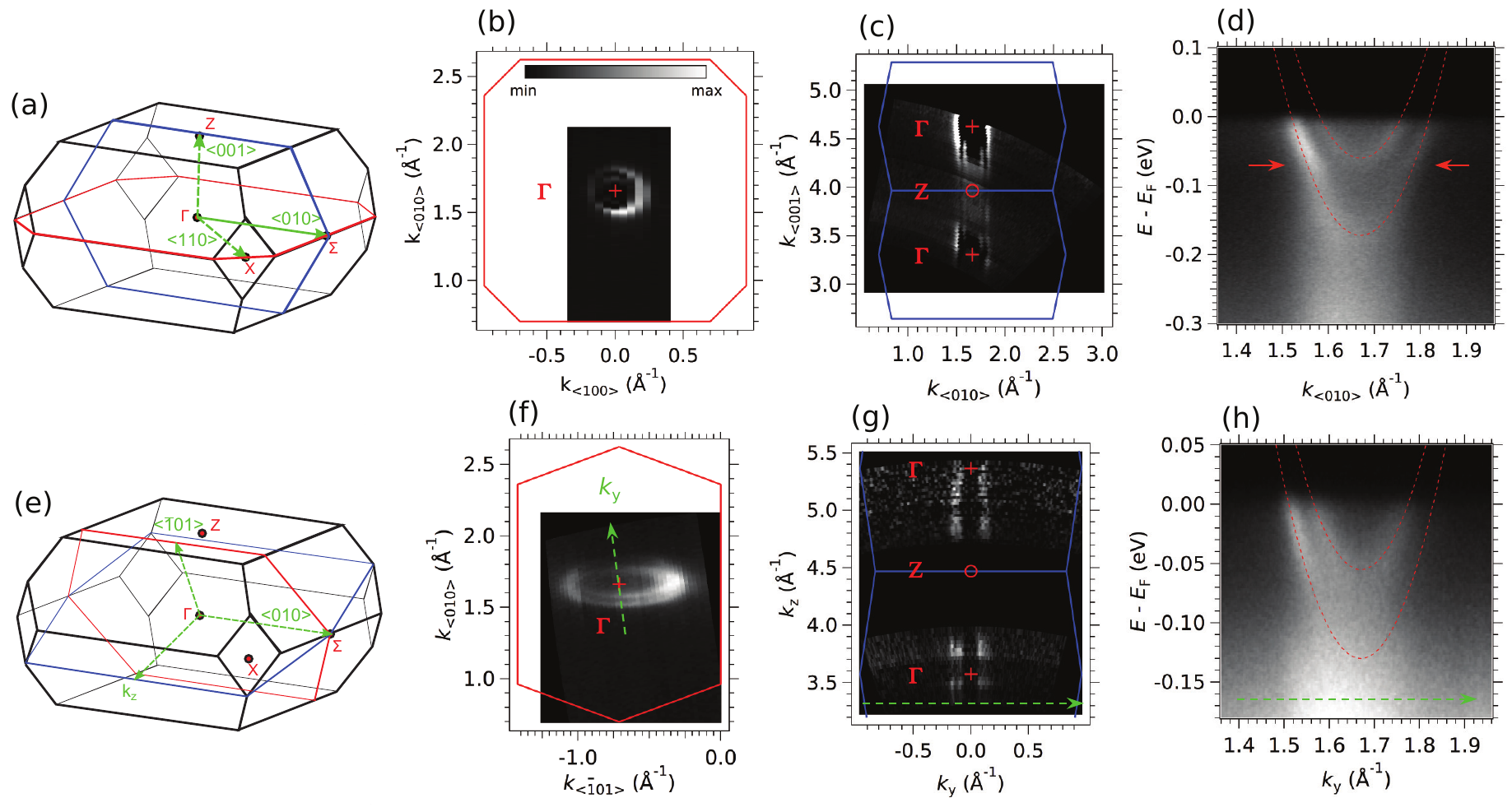}
  \end{center}
  \caption{\label{fig:fig_001_101} \footnotesize{(Color online) 
  		(a) Bulk Brillouin zone of anatase, showing the relevant high symmetry directions (green)
  		and planes (red and blue) of the Fermi surfaces and band dispersion at the $(001)$ surface. 
  		(b) Fermi surface map (second derivative of ARPES intensity, negative values) 
  		measured at $h\nu = 47$~eV on a cleaved, bulk insulating TiO$_2$ anatase $(001)$ surface. 
  		The red lines indicate the edges of the bulk Brillouin zones in the $(001)$ plane 
  		around $\Gamma_{012}$.
  		The same relative color scale indicating minimum (min) to maximum (max)
  		ARPES intensities is used in all color figures in the rest of this work.
  		(c) Fermi surface map (second derivative of ARPES intensity, negative values) 
		in the $k_{\langle 001 \rangle}$~--~$k_{\langle 010 \rangle}$, or $(100)$ plane, 
		acquired by varying the photon energy in 1~eV steps 
		between $h\nu_1 = 38$~eV and $h\nu_2 = 85$~eV. 
		To calculate the momentum perpendicular to the
		surface we set the inner potential to $V_0=13$~eV~\cite{Moser2013,Emori2012}.
        The blue lines are the bulk Brillouin zones containing $\Gamma_{102}$ and $\Gamma_{103}$. 
        (d) Energy-momentum map measured at $h\nu = 47$~eV 
        along the $\langle 010 \rangle$ direction
        on the $(001)$ surface. 
		Dashed red lines are parabolic fits to the dispersing light bands.
		Red arrows show a kink attributed to electron-phonon coupling~\cite{Moser2013}.
		(e-h) Same as (a-d) on the anatase $(101)$ surface. 
		Data in panels (f) and (h) were measured at $h\nu = 47$~eV. 
		The green dashed arrow in (f) indicates the direction of measurements along $k_y$ 
		in (g) and (h),
		which is slightly off $\langle 010 \rangle$.
		The out-of-plane Fermi surface map 
		(second derivative of ARPES intensity, negative values) in (g) was measured 
		by varying the photon energy in 1~eV steps between $h\nu_1 = 33$~eV and $h\nu_2 = 51$~eV 
		and between $h\nu_3 = 76$~eV and $h\nu_4 = 103$~eV. 				
		The Supplemental Material includes figures for
		the raw spectra, Fermi surface maps, and the periodicity of the electronic structure 
		in the $(001)$ and $(101)$ planes.
  		}
  	} 
\end{figure*}

Thereafter, we focus on the electronic structure of the 2DEG 
at the Fermi level, obtained once a stable band-bending has been attained 
after weak exposure to UV irradiation (typically $\sim 20$~min)
-- see Fig.~\ref{fig:anatase_UPS}(c).
Figure~\ref{fig:fig_001_101}(a) shows the bulk Brillouin zone of anatase, 
its high symmetry directions, and the planes of measurements at its cleaved $(001)$ surface.
The Fermi surface at the $(001)$ plane, shown in figure~\ref{fig:fig_001_101}(b),
consists of two circles of $d_{xy}$ orbital character,
confirmed by light polarization dependent measurements 
presented in the Supplemental Material,
similar to the case of SrTiO$_3$~\cite{Santander-Syro2011,Meevasana2011,Plumb2013}.
The quasi-2D character of these two electronic states is strictly demonstrated by the 
Fermi surface map in the $\langle 010 \rangle - \langle 001 \rangle$ plane, 
perpendicular to the surface plane,
shown in figure~\ref{fig:fig_001_101}(c).
This Fermi surface shows that the bands are essentially non dispersing 
along $k_{\langle 001 \rangle}$ over a bulk Brillouin zone, 
thereby confirming the confined character of the electrons close to the surface.
Figure~\ref{fig:fig_001_101}(d) presents the energy-momentum map 
close to the bulk $\Gamma_{102}$ point 
along the cut parallel to the $\langle 010 \rangle$ direction shown in panel (a). 
Two dispersive light bands are visible. 
Their bottoms are at $-60$~meV for the upper band, and $-172$~meV for the lower band. 
The Fermi momenta of $0.09$~\AA$^{-1}$ and $0.16$~\AA$^{-1}$ 
and the experimental band dispersions are given by the peak positions
of the momentum distribution curves (MDCs) at and below $E_F$.
A parabolic fit to the band dispersion 
yields an effective mass of approximately $0.5 m_e$,
which agrees well with theoretical~\cite{Huy2011, Kamisaka2009} 
and experimental~\cite{Hirose2009} results on bulk Nb-doped anatase.
Assuming two spin orientations per band,
this gives an electron concentration of 
$n_{2D}^{(001)} \approx 5.4 \times 10^{13}$~cm$^{-2}$, or about $0.08$ electrons per $a^2$, 
where $a$ is the short lattice constant of the tetragonal lattice 
and $a^2$ the cross section of the unit cell in the $(001)$ plane.
Note that a kink in the dispersion at a binding energy of about $70$~meV is visible (red arrows). 
In the case of bulk states of anatase, such a kink has been attributed to electron-phonon coupling 
and studied in detail by Moser \textit{et al.}~\cite{Moser2013}. 

Next, we present the ARPES measurements on the $(101)$ surface of anatase. 
The bulk Brillouin zone is shown again for clarity in Fig.~\ref{fig:fig_001_101}(e), 
together with relevant directions for the $(101)$ surface and the planes of the
measured Fermi surfaces.
Figure~\ref{fig:fig_001_101}(f) shows the Fermi surface on the $(101)$ plane.
It consists of two ellipses of identical shape, 
corresponding again to $d_{xy}$ orbitals, 
similar to the $d_{xy}$ derived ellipses observed for the 2DEG at the 
SrTiO$_3$~$(110)$ surface~\cite{Wang2014,Roedel2014}.
The Fermi surface map in the $k_{y} - k_{z}$ plane, perpendicular to the cleaved surface, 
is shown in figure~\ref{fig:fig_001_101}(g). The bands are essentially not dispersing 
along $k_{z}$, indicating their 2D character.
Note that, in contrast to the $(001)$ surface, 
the intensity of the 2DEG states in the $k_{z} - k_{y}$ plane
at the $(101)$ surface drops quickly far from bulk $\Gamma$ points, 
due to final-state effects in the photoemission cross-section 
(see the discussion in the Supplemental Material of~\cite{Roedel2014}). 
Figure~\ref{fig:fig_001_101}(h) shows the energy-momentum map 
close to the bulk $\Gamma_{210}$ point along the $k_{y}$ direction. 
As in the $(001)$ surface, two light bands, upper and lower, are also observed in this case. 
Their bottoms of are located at $-60$~meV and $-130$~meV, respectively.
The Fermi momenta taken from the Fermi surface in Fig.~\ref{fig:fig_001_101}(f) 
are $0.41$~\AA$^{-1}~\&~0.33$~\AA$^{-1}$ in the  $\langle \overline{1}01 \rangle$ direction 
and $0.15$~\AA$^{-1}~\&~0.10$~\AA$^{-1}$ in the $\langle 010 \rangle$ . 
The parabolic dispersion based on the experimental Fermi momenta and band bottom energies
yields an effective mass of $m_{<010>}^{(101)}=0.6m_e$ along the $<010>$ direction, 
very close to the one measured at the $(001)$ surface along the same direction.
Assuming again 2 spins per band, this gives an electron concentration of 
$n_{2D}^{(101)} \approx 1.5 \times 10^{14}$~cm$^{-2}$,
three times larger than the one at the $(001)$ surface, 
which is probably related to the higher concentration of oxygen vacancies 
at the $(101)$ surface~\cite{Thomas2007}.

The presence of two non-degenerate bands of identical orbital character directly 
implies that the probed electronic structure is not simply the one expected for bulk anatase.
For SrTiO$_3$, the existence of different subbands has been related to the confinement 
of the electronic states in a quantum well at the surface~\cite{Santander-Syro2011, Meevasana2011}.
More recently, spin-resolved ARPES experiments related the two $d_{xy}$-bands in SrTiO$_3$ 
to different spin polarizations of the ground-state $d_{xy}$ subband~\cite{Santander-Syro2014}. 
Whether or not a magnetic order is also present at the oxygen-deficient surface of TiO$_2$ anatase 
cannot be deduced from the ARPES data presented in this paper. 
Note that magnetic order in thin films of TiO$_2$ due to oxygen vacancies 
was previously observed and discussed~\cite{HoaHong2006,Rumaiz2007,DaeYoon2008,Golmar2008,Coey2010,Coey2010a},
and is thus a potential reason for the observed splitting of the $d_{xy}$-bands.
Further experimental work is necessary to clarify the possible magnetism 
at the oxygen-deficient surface of TiO$_2$ anatase.

In contrast to the 2DEGs at the surface of SrTiO$_3$ and KTaO$_3$, 
our ARPES data in TiO$_2$ anatase 
do not show any heavy ($d_{zx}$ or $d_{yz}$) subbands 
--which we confirmed by complementary measurements as a function of photon polarization and energy,
shown in the Supplemental Material. 
The degeneracy of the three $t_{2g}$ orbitals is lifted in bulk anatase~\cite{Asahi2000} 
due to its tetragonal crystal structure and the inequivalent Ti-O bonding in the $x-y$ plane 
compared to the $z$ direction. 
Consequently, the electron gases confined at the surfaces of TiO$_2$ anatase 
are only composed of electrons of $d_{xy}$-orbital character. 
As shown in the Supplemental Material, 
the spatial extent of the 2DEGs in anatase
is similar to the ones at the surfaces of SrTiO$_3$ and KTaO$_3$, namely about $2$~nm.

\begin{figure}[t]
  \begin{center}
   	  \includegraphics[clip, width=8cm]{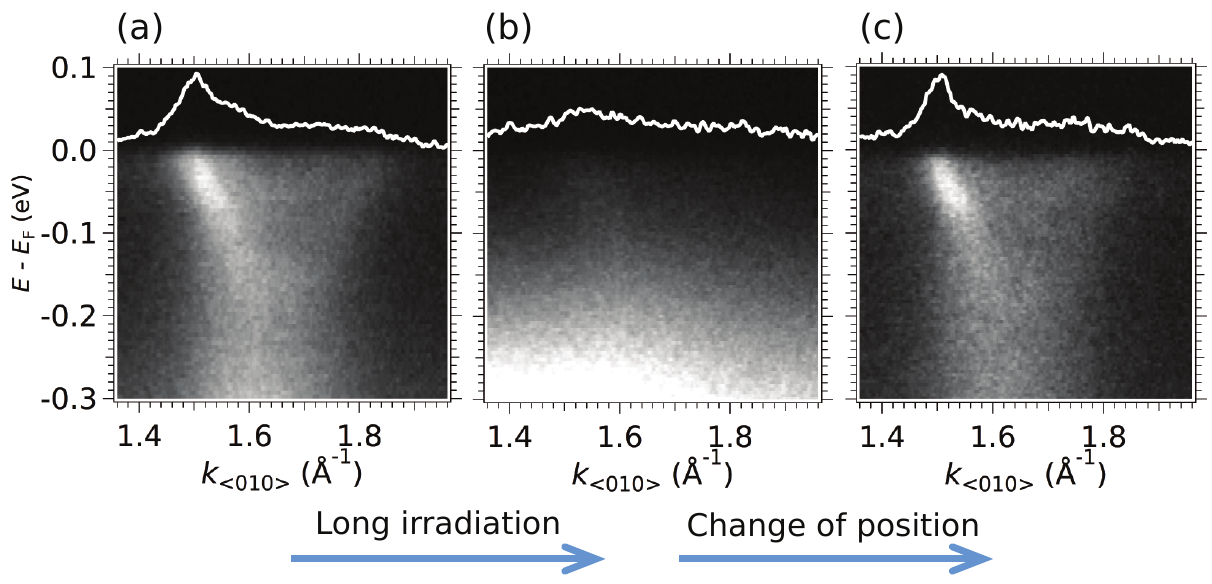}
  \end{center}
  \caption{\label{fig:anatase_UV_destr2DEG} \footnotesize{(Color online) 
      Energy-momentum maps measured at $h\nu = 47$~eV along the $\langle 010 \rangle$ direction
      at the anatase $(001)$ surface for different UV doses.
      (a) After a few minutes of UV irradiation.
      (b) After several hours of exposure to UV light.
      (c) After moving the UV spot to a region in the sample not irradiated before,
      and illuminating again for a few minutes.
      The white curves correspond to the MDCs integrated over $E_F \pm 10$~meV, 
      and are plotted using the same absolute vertical scale in the three panels. 
  	  }
  	} 
\end{figure}

We finally show that exposure to a large UV irradiation dose 
leads to the local destruction of the 2DEG.
Figure~\ref{fig:anatase_UV_destr2DEG}(a) displays the energy-momentum map around $\Gamma_{012}$
measured in a freshly-cleaved anatase $(001)$ surface after only a few minutes of UV irradiation.
The 2DEG is clearly visible. However, as shown in Fig.~\ref{fig:anatase_UV_destr2DEG}(b),
the spectral weight of the dispersing features essentially disappears 
after 1-2 hours of exposure to UV light, 
leaving almost no intensity at the Fermi level.
As seen in figure~\ref{fig:anatase_UV_destr2DEG}(c),
the 2DEG is fully recovered after moving the UV spot to a neighboring region in the sample,
not irradiated before, and illuminating for a few minutes.
As shown in the Supplemental Material,
a high concentration of oxygen vacancies induced by the UV irradiation
also destroys the crystallinity at the surface,
which might in turn lead to the localization of the itinerant electrons
forming the 2DEG.
As the size of the beam spot is only $\approx 50 \times 50$~$\mu\text{m}^2$,
the strong reduction of the surface is local.

In conclusion, we demonstrated the existence of 2DEGs 
at the $(001)$ and $(101)$ surface of TiO$_2$ anatase. 
The two $d_{xy}$ subbands composing the 2DEGs form concentric  
circular Fermi surfaces at the $(001)$ termination 
and ellipsoidal Fermi surfaces at the $(101)$ termination. 
Such orientational tuning of the Fermi sea symmetries is wholly analogous
to the one found at different surfaces of SrTiO$_3$~\cite{Roedel2014}.
Furthermore, we found that light irradiation locally dopes the anatase surfaces,
while the behavior of the excess electrons, \emph{i.e.} delocalized or localized, 
depends on the concentration of oxygen vacancies at the surface, 
eventually resulting in the local destruction of the 2DEG after a high UV irradiation dose. 
This effect, never observed for 2DEGs at other oxide surfaces,
shows that UV light can be used to tailor locally 
the surface order/disorder in anatase, and thus engineer nano-patches 
of metallic 2DEG alongside patches of disordered insulating material.

We thank Bruno Domenichini for discussions.
Work at the CSNSM is supported by public grants from the French National Research Agency (ANR) 
(project LACUNES No ANR-13-BS04-0006-01) 
and the ``Laboratoire d'Excellence Physique Atomes Lumi\`ere Mati\`ere'' (LabEx PALM project ELECTROX) 
overseen by the ANR as part of the ``Investissements d'Avenir'' program (reference: ANR-10-LABX-0039).
T.~C.~R. acknowledges funding from the RTRA--Triangle de la Physique (project PEGASOS).
A.F.S.-S. and M.G. acknowledge support from the Institut Universitaire de France.



\section{SUPPLEMENTAL MATERIAL}

\subsection{Methods for sample preparation and measurements}
The non-doped, polished crystals of TiO$_2$, 
of typical sizes $5\times5\times0.5$~mm$^3$, were supplied by SurfaceNet GmbH. 
The anatase crystals were natural grown and of orange color,
with less than 5~ppm of Mn impurities.
The orientation of the large surface of the crystals was either $(101)$ 
(natural cleaving plane) or $(001)$ 
for anatase, and $(110$) for rutile.
Clean and crystalline surfaces were obtained by cleaving the crystals \emph{in-situ},
using the standard top-post procedure~\cite{Santander-Syro2011},
at temperatures between 7~K and 25~K and pressure lower than $6\times10^{-11}$~Torr. 
To provide electrical grounding to the exposed surface,
the crystals were glued with a conducting silver epoxy to the sample holder
and, after gluing the top-post, wholly covered with graphite paint.
The same protocol has been successfully used in previous ARPES studies of other oxides,
such as SrTiO$_3$ and KTaO$_3$, that are transparent 
and highly insulating in the bulk~\cite{Santander-Syro2011,Santander-Syro2012}.
The sharp photoemission lines and dispersing bands in the ARPES measurements 
demonstrate a crystalline surface. For the different cleaved surfaces, 
the periodicity of the measured electronic structure,
discussed later (figure~\ref{fig:figS25}), 
corresponds to that of an unreconstructed surface, 
and confirms the surface orientation.

The ARPES measurements were conducted at the Synchrotron Radiation Center 
(SRC, University of Wisconsin, Madison, USA) 
and the CASSIOPEE beamline of Synchrotron SOLEIL (Saint-Aubin, France)
We used linearly polarized photons in the energy range $38-103$~eV 
and hemispherical electron analyzers with vertical slits.
The total angle and energy resolutions were $0.25^{\circ}$ and 25~meV at SRC,
and $0.25^{\circ}$ and 15~meV at Soleil. 
The mean diameter of the incident photon beam was smaller than 50~$\mu$m at SOLEIL,
and about 200~$\mu$m at SRC.
The photon flux at SOLEIL was $\sim 2 \times 10^{13}$~photons$/$s$/0.1$bw, 
and about 3 times smaller at SRC. Thus, the density of photons irradiating the sample
was approximately $50$ times larger at SOLEIL than at SRC.
The results were reproduced for at least three different samples for each surface orientation.

\subsection{Rutile vs. anatase}
\begin{figure*}
  \begin{center}
   	  \includegraphics[clip, width=16cm]{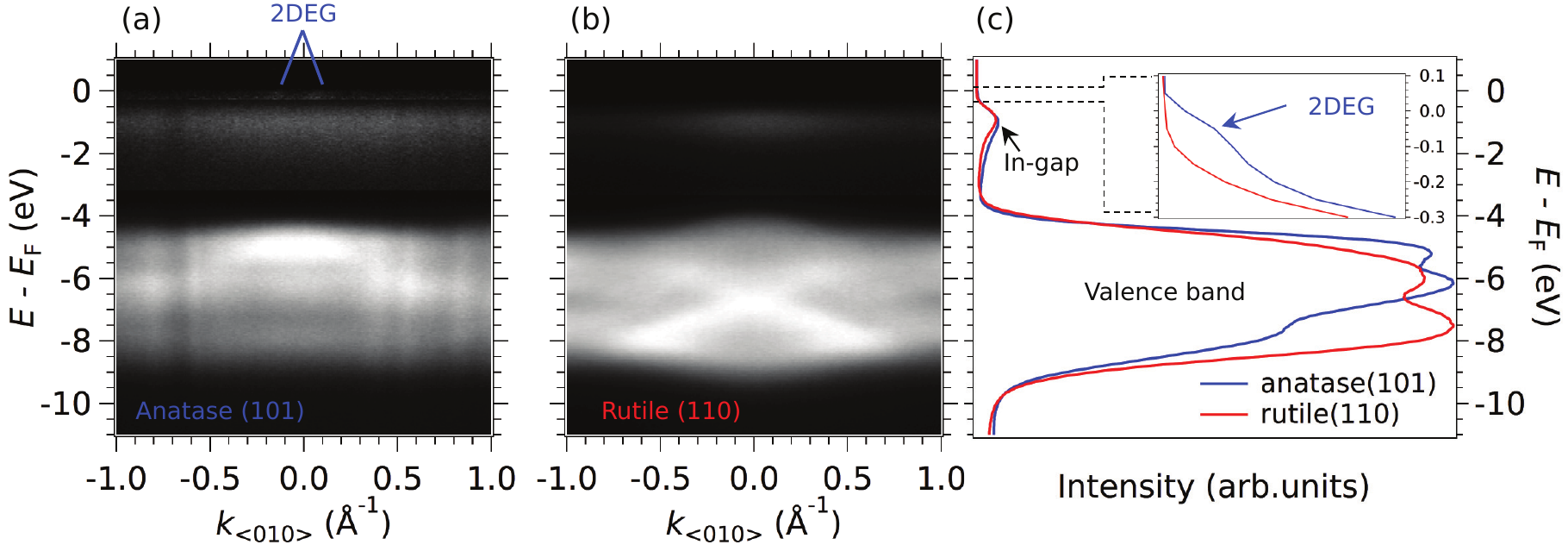}
  \end{center}
  \caption{\label{fig:figS0} \footnotesize{(Color online) 
  	  (a,b)~Energy-momentum maps measured at $h\nu=100$~eV 
  	  along the $\langle 010 \rangle$ or the $\langle 001 \rangle$ direction 
  	  at the anatase $(101)$ and rutile $(110)$ surface. 
  	  The intensities in different energy intervals (valence band, in-gap, 2DEG) 
  	  were normalized to make all the features visible. 
  	  (c) Angle-integrated, raw spectra of the two energy-momentum maps in (a,b). 
  	  The inset shows a zoom of the angle-integrated spectra (log scale)
  	  for angles close to normal emission. 
  	  The spectral weight at the Fermi level in the case of the anatase $(101)$ surface 
  	  corresponds to the 2DEG.
  	  }
  	} 
\end{figure*}

The different behavior of excess electrons induced by
oxygen vacancies in rutile and anatase was already discussed
in the main text. In contrast to anatase, the excess electrons in rutile do not 
fill up the conduction band (forming the 2DEG), but are localized at lattice Ti
sites~\cite{Setvin2014}. Accordingly, we did not detect any dispersing features at
the Fermi level at the surface of rutile $(110)$, neither for cleaved nor for \emph{in situ}
prepared surfaces under various measurement conditions (photon energy, emission
angle, light polarization) at $T = 7$~K.
To illustrate the difference between the two systems, 
figure~\ref{fig:figS0}(a,b) shows the valence band, the in-gap state corresponding
to the localized electrons, and in the case of anatase the 2DEG measured at normal emission 
at $h\nu=100$~eV. The angle-integrated data in figure~\ref{fig:figS0}(c)
demonstrates that the concentration of oxygen vacancies is similar in the two cases, 
as the in-gap states have similar intensities. 
The failure to observe dispersing features at the Fermi level of rutile 
shows that oxygen vacancies at the surface are not sufficient to create a 2DEG, 
and that structural factors play a crucial role regarding the
localization/delocalization of excess electrons.

\subsection{Surface vs. bulk-like electronic structure of anatase}
Moser \textit{et al.}~\cite{Moser2013} characterized the polaronic conduction 
for different charge carrier densities (controlled by the oxygen partial pressure sample chamber) 
at the anatase TiO$_2$(001) surface using ARPES.
In the present work, we only show data after reaching the saturation value 
of the charge carrier density.
Intermediate values of the charge carrier density can be observed 
but are not stable long enough to measure with a sufficient signal-to-noise ratio 
without controlling the oxygen partial pressure in the measurement chamber.
Contrary to our results, Moser \textit{et al.} observed only one dispersing subband 
of 3D character (\emph{i.e.}, dispersing along the direction perpendicular to the surface) 
at low charge carrier densities.
Note that for a low concentration of oxygen vacancies and charge carriers, 
the potential well at the surface is rather shallow and the electrons are thus barely confined. 
But, even for charge carrier densities comparable to our results 
($n_{3D,\text{Moser}}\approx 2\times10^{20}\text{~cm}^{-3}$ 
to $n_{3D}\approx (n_{2D}^{(001)})^{(3/2)}\approx 4\times10^{20}\text{~cm}^{-2}$) 
Moser \textit{et al.} observed only one dispersing band. 
Such discrepancy in the number of bands and dimensionality occurred already in the case of 
the electronic structure at the surface of SrTiO$_3$: Chang \textit{et al.}~\cite{Chang2010}, 
using the same ARPES setup as Moser \textit{et al.}, 
measured only one dispersing band of $d_{xy}$ character and attributed it to the bulk, 
whereas two $d_{xy}$ bands and 2D behavior were observed 
in other works~\cite{Santander-Syro2011, Meevasana2011, Plumb2013}.
Both studies observing only one $d_{xy}$ band were conducted 
at the same synchrotron endstation at the Advanced Light Source. 
Thus, the discrepancy in the number of bands might be related to the specifics of the ARPES endstation. 
Possible reasons are a lower photon flux or different photoemission selection rules
due to different measurement geometries. 
For instance, for SrTiO$_3$, we checked (not shown)
that using horizontal slits instead of vertical slits for the electron analyzer yields
only one of the two $d_{xy}$ subbands. 

\subsection{Spatial extent of the 2DEG at the surface of anatase}
An estimate of the spatial extent of the 2DEGs in anatase using the triangular potential well model 
used previously~\cite{Santander-Syro2011} can be obtained, subject to one caveat. 
If the subband splitting is due to the quantum confinement, the width of the  
potential well follows directly from this splitting and the effective mass of the subbands
along the confinement direction~\cite{Santander-Syro2011}. 
For anatase, $m_{eff}^{\langle 001 \rangle} = 4.65 m_e$~\cite{Kamisaka2009}, 
yielding a confinement length of $1.8$~nm for the 2DEG at the anatase $(001)$ surface.
If, on the other hand, the subband splitting is due to magnetism~\cite{Santander-Syro2014}, 
one can assume that the next quantum-well states of $d_{xy}$ character
lie just above $E_F$, which gives an upper bound of $1.4$~nm for the width of the 
potential well at the $(001)$ surface.
One sees that, in any case, the confinement length of the 2DEGs at the surface of anatase 
is similar to the ones at the surfaces of SrTiO$_3$ and KTaO$_3$, of about $2$~nm.

\subsection{Influence of UV synchrotron irradiation}
\begin{figure*}
  \begin{center}
   	  \includegraphics[clip, width=16cm]{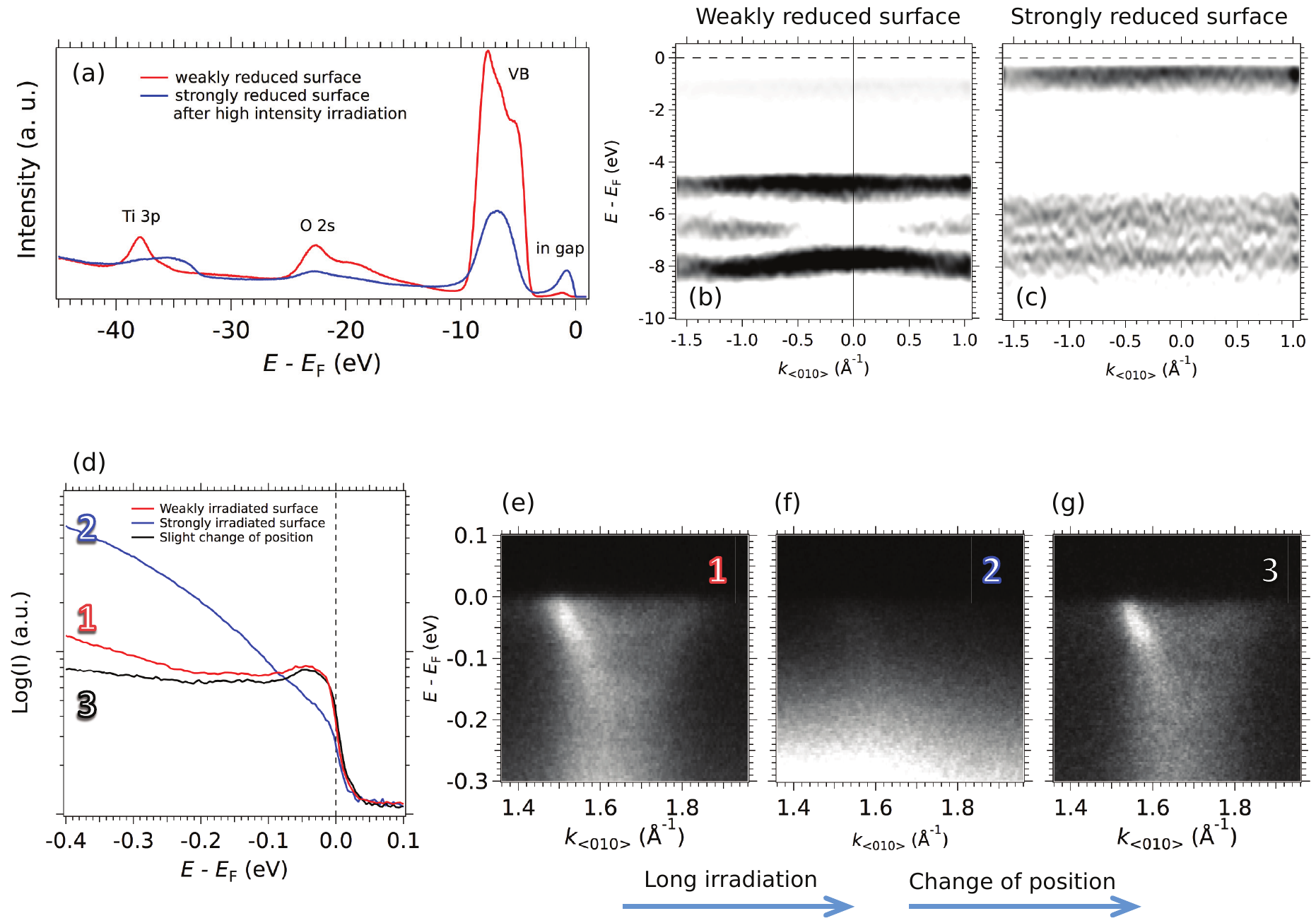}
  \end{center}
  \caption{\label{fig:figS1} \footnotesize{(Color online) 
  	   (a) Angle-integrated spectra of a cleaved TiO$_2$(001) anatase sample measured 
  	   at $h\nu=100$~eV showing the density of states for binding energies 
  	   between $-45$~eV and $2$~eV. The red curve corresponds to a weakly reduced surface, 
  	   the blue one to a surface exposed to the intense zero-order light irradiation of the synchrotron. 
       (b,~c) Corresponding angle-resolved energy-momentum maps 
       (second energy derivatives, negative values only) zoomed over the valence-band region. 
       While the valence band is dispersive in the crystalline, weakly irradiated surface, 
       it becomes non-dispersive and featureless in the amorphous, strongly irradiated surface.
       (d) Angle-integrated spectra measured at $h\nu = 47$~eV (log scale for spectral intensity), 
       showing the tail of the in-gap peak and the intensity at the Fermi level. 
       The red curve (label 1) corresponds to a freshly cleaved sample irradiated for a first sweep, 
       the blue curve (label 2) to a surface exposed to irradiation for several hours,
       and the black curve (label 3) to the electronic structure after changing the sample position.
       (e,~f,~g) Energy-momentum maps measured at $h\nu = 47$~eV along the $\langle 010 \rangle$ direction.  
       The momentum-integrated curves of these maps corresponds to the data shown in (b) 
       as indicated by the labels 1, 2 and 3.
       }
  	} 
\end{figure*}
We present now supplementary data regarding the influence of the UV synchrotron irradiation 
on the stoichiometry and electronic structure at the surfaces of anatase. 

We conducted our measurements at two different synchrotrons: 
SRC ($2^{nd}$ generation) and Soleil ($3^{d}$ generation). 
The normal measurement conditions (photon flux density at the surface of the sample, 
vacuum pressure, temperature) at SRC enable to reduce the surface, 
create the 2DEG, and study a stable 2DEG for days. 
In contrast, the much higher photon flux density at Soleil creates so many oxygen vacancies 
that the induced disorder at the surface lead to the disappearance of the 2DEG 
within about one hour at $T=7$~K at the anatase $(101)$ surface. 
As this effect was never observed for the 2DEG at the surface of SrTiO$_3$ and KTaO$_3$, 
we conclude that the anatase surfaces are much more sensible to the UV irradiation.

The effects of the synchrotron UV light on the $(001)$ surface are shown in figure~\ref{fig:figS1}.
An approximate upper limit for the effects is shown in Fig.~\ref{fig:figS1}(a),
where we present the angle-integrated spectra measured at $h\nu=100$~eV 
on a fractured $(001)$ anatase sample for binding energies between $-45$~eV and $2$~eV. 
The red curve corresponds to a weakly reduced surface, similar to the data in figure~1
in the main text, and the blue curve to the electronic structure of the surface 
after exposing the sample to high intensity zero-order irradiation of the synchrotron.
As can be seen from figure~\ref{fig:figS1}, there are several changes in the electronic structure
induced by the severe exposure to UV light. 
First, the shoulder of the Ti-$3p$ peak, corresponding to Ti$^{3+}$ states, transforms 
to become the the main Ti-$3p$ peak at even lower binding energies, 
indicating that the majority of Ti atoms close to the surface have a low valency 
$(\leq 3+)$ due to the creation of many oxygen vacancies. 
Second, the shape of the valence band becomes featureless 
and its dispersion disappears after a high UV dose, 
demonstrating the large disorder induced at the surface. 
Third, a high concentration of oxygen vacancies is obvious from the increased intensity 
of the in-gap peak and the decreased intensity of the peaks related to oxygen 
(O-$2s$ and valence band). 
These measurements clearly demonstrate that the surface structure can be altered drastically,
to the point of completely destroying the crystalline order, 
due to light irradiation.

Figures~\ref{fig:figS1}(b,~c) show the angle-resolved measurements corresponding to the spectra
in figure~\ref{fig:figS1}(a), zooming over the valence-band region. 
The clear dispersion of the valence band in the weakly reduced surface, Fig.~\ref{fig:figS1}(b), 
is a direct consequence of the good surface crystallinity (Bloch theorem). 
After intense UV irradiation, Fig.~\ref{fig:figS1}(c), the surface crystallinity is destroyed, 
and the valence-band becomes non-dispersive and featureless. 

Even without the zero-order irradiation, the reduction of the surface 
can lead to the disappearance of the 2DEG. 
The angle-integrated spectra in figure~\ref{fig:figS1}(b) were measured at $h\nu = 47$~eV 
and show the tail of the in-gap peak and the intensity at the Fermi level 
for binding energies between $-0.4$~eV and $0.1$~eV. 
After several hours of irradiation (blue curve), the intensity of the in gap state increases 
and the intensity at the Fermi level corresponding to the 2DEG decreases. 
The state of the weakly irradiated surface (red curve) 
can be retrieved by slightly changing the position of the UV light spot 
on the sample surface (black curve). 
As the size of the beam spot is only $\approx 50 \times 50$~$\mu\text{m}^2$, 
the strong reduction of the surface is local. 
This results suggests that UV light can be used to tailor locally 
the surface order/disorder in anatase,
and thus write mesoscopic or nanoscopic patches of metallic 2DEG intercalated with patches
of disordered insulating material.

The energy momentum maps corresponding to the angle-integrated spectra of figure~\ref{fig:figS1}(b)
were already shown and discussed in figure~3 of the main text.
We display them again, for completeness, in figures~\ref{fig:figS1}(c,d,e). 
The 2DEG is clearly visible in figure~\ref{fig:figS1}(c), 
whereas there is almost no intensity at the Fermi level in figure~\ref{fig:figS1}(d) 
at the strongly reduced surface, and the 2DEG is recovered in figure~\ref{fig:figS1}(e),
after moving the UV spot to a neighboring region in the sample.

\begin{figure}[t]
  \begin{center}
   	  \includegraphics[clip, width=9cm]{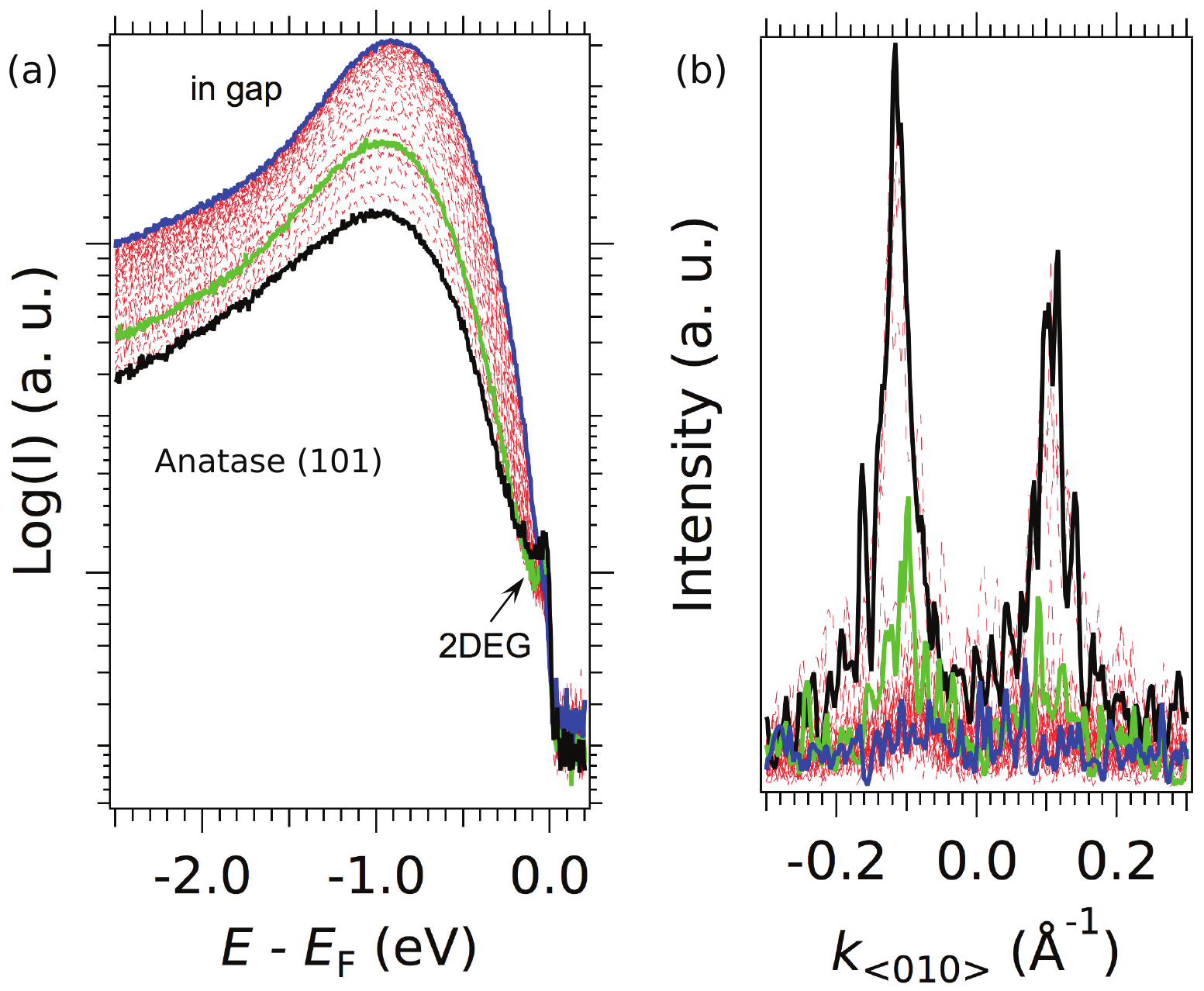}
  \end{center}
  \caption{\label{fig:figS2} \footnotesize{(Color online) 
  		(a) Angle-integrated spectra (log scale for intensities) measured at equal time intervals 
  		at $h\nu = 47$~eV for binding energies between $-2.5$~eV and $0.2$~eV.
  		One observes the in-gap peak and the intensity of the 2DEG at the Fermi level. 
  		Three of the curves are highlighted in black,  green and blue corresponding to the start, 
  		a midpoint and the end of these measurements.
  		(b) Corresponding momentum distribution curves at the Fermi level. 
  		}
  	} 
\end{figure}

Another interesting fact is that the anatase $(101)$ surface 
degrades more rapidly than the $(001)$ surface. 
The data corresponding to this degradation are shown in figure~\ref{fig:figS2}. 
The angle-integrated spectra (red curves) in Fig.~\ref{fig:figS2}(a) 
were measured at equal time intervals (of approximately 2.5 minutes) 
at a photon energy $h\nu = 47$~eV,  
and show the increase of the intensity of the in-gap peak and decrease 
in the intensity at the Fermi level for binding energies between $-2.5$~eV and $0.2$~eV. 
Three of the curves highlighted in black,  green and blue correspond to the start, 
a midpoint and the end of these measurements. 
The intensity of the 2DEG is evident in the momentum distribution curves at the Fermi level 
in figure~\ref{fig:figS2}(b). At the beginning (black curve) 
two peaks, corresponding to the dispersing bands shown in figure~2(h) of the main text, are visible. 
Their intensity decays rapidly after several minutes, is barely visible (green curve) 
and disappears completely (blue curve). 
Note that the Fermi momenta in the MDCs does not increase 
--compare black and green curve in Fig.~\ref{fig:figS2}(b), 
although more oxygen vacancies are created as evidenced 
by the increase in intensity of the in-gap state in Fig.~\ref{fig:figS2}(a).
As the electron density in the conduction band is proportional to the Fermi momenta $n\propto k_F^2$, 
the excess electrons due to oxygen vacancies seem to populate the in-gap state (localized electrons) 
and not any longer the conduction band (delocalized electrons). 
This observation is in agreement with the observed saturation of the band bending in the main text.

A high concentration of oxygen vacancies induced by the synchrotron irradiation
creates a high degree of disorder at the surface. 
If the degree of disorder at the surface is sufficiently high, no dispersing bands exist 
as there is no longer the periodicity of the lattice, and the 2DEG disappears, 
as effectively seen in Fig.~\ref{fig:figS2}(b).
            
\subsection{Raw data and periodicity of the electronic structure}
\begin{figure*}
  \begin{center}
   	  \includegraphics[clip, width=16cm]{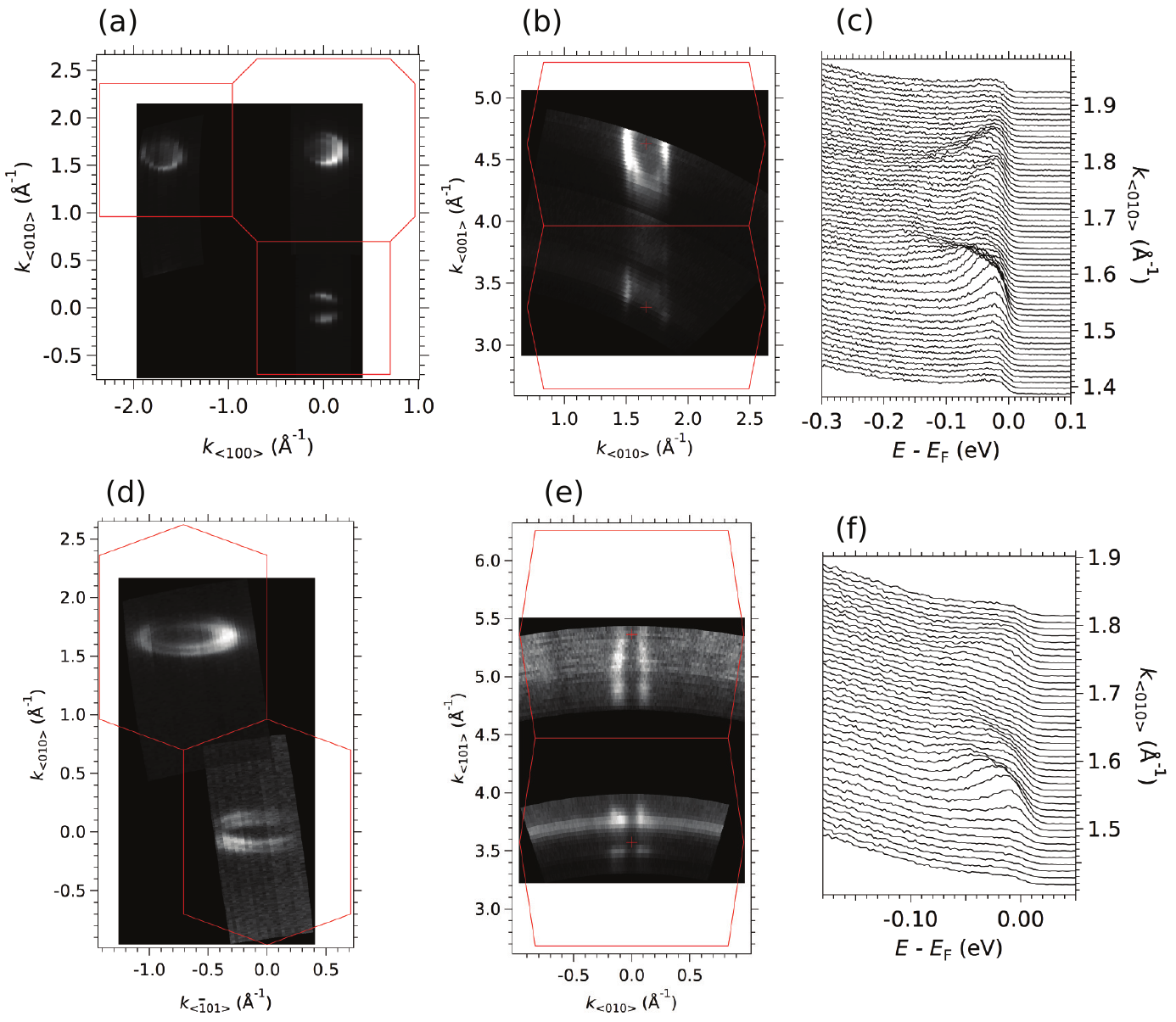}
  \end{center}
  \caption{\label{fig:figS25} \footnotesize{(Color online) 
  		(a) Raw Fermi surface map measured at $h\nu = 47$~eV on a cleaved, 
  		bulk insulating TiO$_2$ anatase $(001)$ surface. 
  		The red lines indicate the edges of the bulk Brillouin zones
  		in the $(001)$ plane of $\Gamma_{102}$.
  		(b) Raw Fermi surface map in the $k_{\langle 001 \rangle}$~--~$k_{\langle 010 \rangle}$, 
  		or $(100)$ plane. The red lines are the bulk Brillouin zones containing 
  		$\Gamma_{102}$ and $\Gamma_{103}$. 
  		(c) Raw EDCs measured at $h\nu = 47$~eV ($k_z = 3.45$~\AA$^{-1}$)
   		along the $\langle 010 \rangle$ direction on the $(001)$ surface. 
  		They correspond to the energy-momentum map shown in Fig.~2(d) of the main text.
  		(d) Raw Fermi surface map measured at $h\nu = 47$~eV on a cleaved, 
  		bulk insulating TiO$_2$ anatase $(101)$ surface. 
  		The red lines indicate the edges of the unreconstructed Brillouin zones 
  		in the $(101)$ plane of $\Gamma_{002}$. 
		(e) Raw Fermi surface map in the $k_{y} - k_{z}$ plane 
		measured on a cleaved, bulk insulating TiO$_2$ anatase $(101)$ surface. 
		The red lines are the bulk Brillouin zones containing $\Gamma_{200}$ and $\Gamma_{300}$. 
  		(f) Raw EDCs measured at $h\nu = 47$~eV ($k_z = 3.45$~\AA$^{-1}$)
  		along the $\langle 010 \rangle$ direction on the $(101)$ surface. 
  		They correspond to the energy-momentum map shown in Fig.~2(h) of the main text.}
  	} 
\end{figure*}

Figure~\ref{fig:figS25} shows the raw Fermi surface maps 
and energy distribution curves (EDCs) of the data presented in Fig.~2 of the main text. 
Comparing the raw maps of Figs.~\ref{fig:figS25}(a,~b,~d,~f)
with the second-derivative maps of Figs.~2(b,~c,~f,~g) in the main text,
it is evident that the second derivative in the Fermi surface maps
only enhances the peak-to-background ratio,
and does not create artifacts in the intensity distribution. 

Additionally, the Fermi surface maps in figures~\ref{fig:figS25}(a,~d), 
which span a portion of in-plane momentum space larger than in Fig.~2 of the main text, 
show that the periodicity of the electronic structure 
at the cleaved $(001)$ and $(101)$ surfaces, respectively,
corresponds to the one expected at unreconstructed surfaces. 

Note that the Fermi surface map of Figure~\ref{fig:figS25}(b),
and the corresponding dispersions at $h\nu = 47$~eV ($k_z = 3.45$~\AA$^{-1}$) 
shown in Fig.~3 of the main text, 
were not obtained at normal emission, but in the second Brillouin zone. 
Thus, the mere geometry of measurements imposes already extrinsic intensity asymmetries 
between the the left and right branches of the 2DEG electron bands.

\subsection{Polarization dependence}
\begin{figure*}
  \begin{center}
   	  \includegraphics[clip, width=16cm]{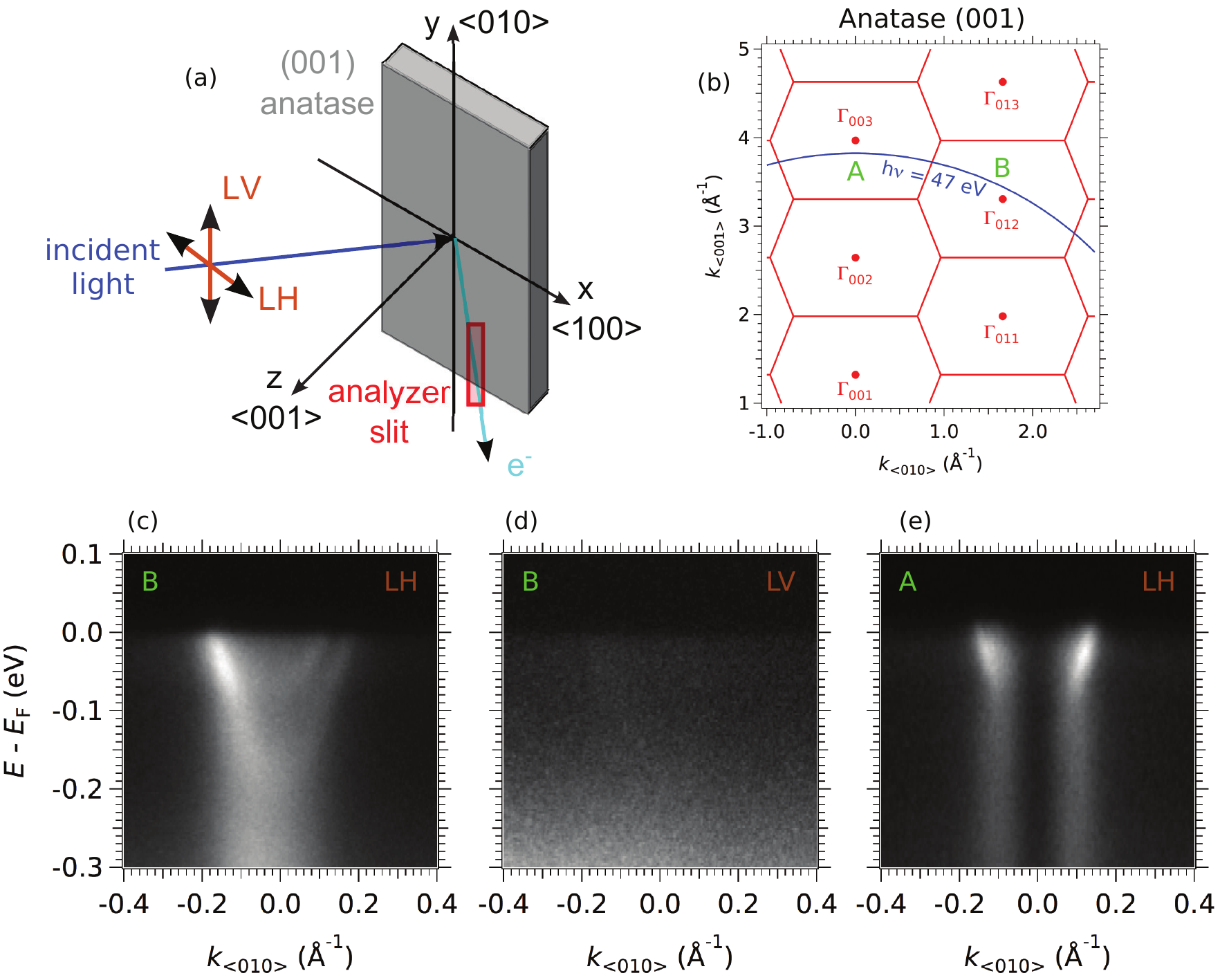}
  \end{center}
  \caption{\label{fig:figS4} \footnotesize{(Color online) 
  		(a) Sketch of the experimental ARPES setup.
		(b) Schematic representation of the reciprocal space of bulk TiO$_2$ anatase 
		over several Brillouin zones in the $(100)$ plane.
		(c) Energy momentum map measured at $h\nu=47$~eV with linear horizontal light polarisation (LH) 
		at the surface of a tilted sample corresponding to position B in reciprocal space in (b).
		(d) Same as (c) but with linear vertical light polarisation (LV).
		(e) Same as (c) but at normal emission, position A in (b).
  		}
  	} 
\end{figure*}

The polarization dependence of the $t_{2g}$ orbitals ($d_{xy}$,$d_{xz}$,$d_{yz}$) 
forming the 2DEG at the $(001)$ surface of SrTiO$_3$ was discussed 
in previous works~\cite{Santander-Syro2011,Yukawa2013} and can be directly applied 
to the $(001)$ surface of anatase. 
To understand the photoemission selection rules, the geometry of the experimental setup is essential 
and shown in Fig.~\ref{fig:figS4}(a). We used light polarized horizontally (LH, in the $x-z$ plane) 
and vertically (LV, along the $y$-direction). 
The slits of the analyzer are aligned vertically along the $y$-direction
and the sample was rotated around the $x$ axis.
In this geometry, the measurement plane containing 
the electrons ejected from the surface and entering the detector through the vertical slits,
coincides with the $yz$ plane. 

The measurements were conducted at a photon energy of $h\nu=47$eV both at normal emission
and tilted by $26^\circ$ around the $x$-axis. 
The corresponding positions in reciprocal space of bulk anatase, labelled A and B, 
are shown in figure~\ref{fig:figS4}(b).

To measure a non-zero intensity, the final state needs to be even 
with respect to the $y-z$ measurement plane. 
As LV polarisation is even with respect to this plane, 
only states with an orbital character even 
to the $y-z$ plane, hence $d_{yz}$ states, can be detected using LV. 
The LH polarisation has even as well as odd components 
with respect to the $y-z$ plane,
and states of all $t_{2g}$ orbitals have non-zero matrix elements using this polarisation. 

The ARPES spectra in figures~\ref{fig:figS4}(c,~d) show the measured intensities 
at the point B in reciprocal space for LH and LV polarizations, respectively. 
The almost zero intensity in LV polarisation demonstrates that the orbital character is not $d_{yz}$.
To exclude one more orbital character we turn to the measured intensities at normal emission (point A) 
in Fig.~\ref{fig:figS4}(e). To have a non-zero matrix element 
at the $\Gamma$ point in normal emission,
the final state has to be even with respect to both the $x-z$ and $y-z$ planes. 
Hence, as the LH polarization is even with respect to the $x-z$ plane, 
only states with an orbital character even to the $x-z$ plane ($d_{xz}$) can be detected. 
The measured intensity at and close to normal emission in figure~\ref{fig:figS4}(e),
\emph{i.e.} around $k=0$, is zero and thus, 
the bands are not of $d_{xz}$ character.

Consequently, the orbital character of the bands is $d_{xy}$, 
in agreement with the circular Fermi surfaces in figure~2(b) of the main text 
and the non-degeneracy of the $t_{2g}$ orbitals in the bulk of anatase. 
We verified the matrix elements at a different photon energy to affirm that the matrix elements 
are due to orbital character and not photon energy.

\end{document}